\documentclass[aps,prb,twocolumn,groupedaddress,amssymb,amsmath,superscriptaddress,floatfix]{revtex4-1}

\usepackage{hyperref}
\usepackage[pdftex]{graphicx}
\usepackage{xr}
\usepackage{bm}      
\usepackage{color}

\newcommand{\ybcoy}{Y\-Ba$_2$\-Cu$_3$\-O$_{6+y}$}
\newcommand{\eubcoy}{Eu\-Ba$_2$\-Cu$_3$\-O$_{6+y}$}

\newcommand{\ybco}{Y\-Ba$_2$\-Cu$_3$\-O$_{6.35}$}
\newcommand{\ybcol}{Y\-Ba$_2$\-Cu$_3$\-O$_{6.23}$}
\newcommand{\eubco}{Eu\-Ba$_2$\-Cu$_3$\-O$_{6.35}$}
\newcommand{\ybcod}{Y\-Ba$_2$\-Cu$_3$\-O$_{6.325}$}
\newcommand{\eubcod}{Eu\-Ba$_2$\-Cu$_3$\-O$_{6.325}$}
\newcommand{\calabalacuo}{Ca$_{x}$La$_{1.25}$Ba$_{1.75-x}$Cu$_3$O$_{6+y}$}

\newcommand{\ycabcoy}{Y$_{1-x}$Ca$_x$\-Ba$_2$\-Cu$_3$\-O$_{6+y}$}

\newcommand{\yeubco}{Y$_{z}$Eu$_{1-z}$\-Ba$_2$\-Cu$_3$\-O$_{6+y}$}
\newcommand{\yRbco}{Y$_{z}$Re$_{1-z}$\-Ba$_2$\-Cu$_3$\-O$_{6+y}$}

\newcommand{\lsco}{La$_{2-x}$Sr$_x$\-CuO$_4$}

\newcommand{\msr}{$\mu$SR}
\newcommand{\yebco}{Y$_{z}$Eu$_{1-z}$\-Ba$_2$\-Cu$_3$\-O$_{6.35}$}

\newcommand*{\unit}[1]{\ensuremath{\mathrm{\,#1}}}
\definecolor{OliveGreen}{rgb}{0,0.5,0}
\definecolor{red70}{rgb}{0.7,0,0}
\definecolor{blu70}{rgb}{0,0,0.7}

\begin{document}


\title{Singling out the effect of quenched disorder in the phase diagram of cuprates.}


\author{R.~De Renzi }
\email[]{roberto.derenzi@unipr.it}
\affiliation{Department of Mathematical, Physical and Computer Sciences, University of Parma, Parco delle Scienze 7/a, 43124 Parma, Italy}

\author{F.~Coneri}
\affiliation{MESA+ Institute for Nanotechnology, University of Twente, 7500 AE Enschede, The Netherlands}

\author{F.~Mezzadri}
\affiliation{Department of Chemistry, Life Science and Environmental Sustainability, University of Parma, Parco delle Scienze 17/A, 43124 Parma, Italy}

\author{G.~Allodi}
\affiliation{Department of Mathematical, Physical and Computer Sciences, University of Parma, Parco delle Scienze 7/a, 43124 Parma, Italy}

\author{G.~Calestani}
\author{L. Righi}
\affiliation{Department of Chemistry, Life Science and Environmental Sustainability, University of Parma, Parco delle Scienze 17/A, 43124 Parma, Italy}
\affiliation{IMEM-CNR, Parco delle Scienze 37/A, 43124 Parma, Italy}

\author{G. M. Lopez}
\author{V. Fiorentini}
\author{A. Filippetti}
\affiliation{Dipartimento di Fisica and CNR-IOM, Universit\`a di Cagliari, 09042 Monserrato, Italy}

\author{S.~Sanna}
\email[]{samuele.sanna@unipv.it}
\affiliation{Department of Physics and Astronomy, University of Bologna and CNR-SPIN, via Berti Pichat 6-2, I-40127 Bologna, Italy}



\date{\today}

\begin{abstract}
  We investigate the specific influence of structural disorder on the suppression of antiferromagnetic order and on the emergence of cuprate superconductivity. We single out pure disorder, by focusing on a series of \yeubco\ samples  at \emph{fixed} oxygen content $y=0.35$, in the range $0\le z\le 1$. The gradual Y/Eu isovalent substitution smoothly drives the system through the Mott-insulator to superconductor transition from a full antiferromagnet with N\'eel transition $T_N=320$ K at $z=0$ to a bulk superconductor with superconducting critical temperature $T_c=18$ K at $z=1$, \ybco. The electronic properties are finely tuned by gradual lattice deformations induced by the different cationic radii of the two lanthanides, inducing a continuous change of the basal Cu(1)-O chain length, as well as a controlled amount of disorder in the active Cu(2)O$_2$ bilayers. We check that internal charge transfer from the basal to the active plane is entirely responsible for the doping of the latter and  we show that superconductivity emerges with orthorhombicity. By comparing transition temperatures with those of the isoelectronic clean system we determine the influence of pure structural disorder connected with the Y/Eu alloy.

\end{abstract}

\maketitle
Disorder has an important role in the physics of cuprates, since it induces nucleation of the  Anderson localization transition from the non-Fermi-liquid superconductor phase to the antiferromagnetic (AF) ordered charge transfer insulator phase. Very strong coupling has been directly investigated by means of Cu substitutional impurities, often employed also directly as a probe. \cite {AlloulRevModPhys2009} An inevitable source of weaker disorder is due to the randomly localized ionic charges and the accompanying local structural distortion in the buffer layer that provides {\em chemical doping}, controlling the electronic transition. It is often referred to as {\em quenched} disorder. In this respect the two most investigated cuprate oxides \lsco\ and \ybcoy\ differ in the distance between the buffer and the active layers. In \lsco\ the doping is tuned by heterovalent La/Sr cation substitution, close to Cu. In \ybcoy, considered the closest case to the  clean limit cuprate for its low degree of quenched disorder and high  optimal $T_c$, two inequivalent Cu ions are present and doping is controlled by oxygen stoichiometry in a basal buffer layer, containing  fragments of Cu(1)O chains (hereafter, chains), whereas magnetism and superconductivity take place in more distant Cu(2)O$_2$ bilayers (hereafter planes).

Theoretically, disorder was addressed explicitly by Alvarez et al. \cite{AlvarezPRB2005}, in a phenomenological model of itinerant electrons on a Cu(2)O$_2$ square lattice showing that disorder does not lead to a universal behavior. For instance towards the clean limit, as in \ybcoy, it may give rise e.g. to coexistence of AF and superconducting order, \cite{SannaPRL2004} or to an instability against charge density waves formation, \cite {Ghiringhelli2012,Wu2015} whereas quenched disorder opens a hole-density {\em window} where none of the two competing orders dominates, both having robust short range correlations. A review by Dagotto\cite{DagottoScience2005} highlighted that complex behaviour may emerge, leveraging on the many available degrees of freedom. Several of  these considerations are in agreement with experimental results\cite{SannaPRL2004,SannaPRBR2010}. However it is experimentally difficult to determine independently the effect of doping and quenched disorder,  \cite{AlloulRevModPhys2009} since they are strictly proportional to one another in all quoted examples. Only \ycabcoy,  to date, allowed this distinction\cite{Uemura1989,Bernhard1996,Awana1996}, thanks to its double doping channel. 

The connection between doping and disorder is drastically altered in \yeubco\ at constant oxygen content $y$, where Y$^{3+}$ is randomly replaced by the isovalent Eu$^{3+}$ cation with a larger radius. The stronger quenched disorder due to nearest neighbor heterovalent cations is absent, leaving only a residual, weakest quenched disorder. However, the distinct cationic radii of Y (101.9 pm) and Eu (106.6 pm)\cite{ShannonRadii} very strongly affect the {\em internal} chemical doping from the buffer to the active layers: \ybcoy\ and \eubcoy\ have very different onset thresholds for superconductivity\cite{LuetgemeierPC1996} ($y>0.40$ for Eu and $y<0.30$ for Y), hence a narrow range of $y$ exists for which \eubcoy\ is still a CT insulating antiferromagnet, whilst \ybcoy\ is a bulk superconductor, despite their nominally equal total cationic charge. This is a well known fact that was never explicitly exploited until now. 

The strong influence of the Y/Eu steric hindrance on electronic properties was convincingly attributed by nuclear quadrupole resonance (NQR) studies\cite{LuetgemeierPC1996}  to the reduction of the average length $\ell$ of the chain fragments. The chain length in turns determines the cation ratio Cu(1)$^{1+}$/Cu(1)$^{2+}$ in the basal layer, hence the doping of the planes. These conclusions are supported by our Density Functional Theory (DFT) computations for a few relevant stoichiometries.

Here, choosing $y=0.35$, we show by susceptibility and muon spin spectroscopy (\msr) measurements that \yebco\ spans the whole phase diagram, from a nearly optimal AF N\'eel state (with $T_N=320(15)$ K, $\mathbf k = (1,1,0)$ and staggered spin basis along c in the bilayer) to a bulk underdoped superconductor with $T_c=20(2)$ K. This means that the electronic properties of the planes are very effectively modulated from those of a charge transfer insulator, across the metal insulator transition, by a mere structural distortion, without altering the total charge count. We show that the plane hole doping is determined by internal charge transfer, modulated by the Cu(1)O$_y$ chain fragment length. 

Although the isoelectronic cation substitution has no influence on the total charge count, it introduces a form of quenched disorder, absent in the end members, $z=0$ and $z=1$, and maximal for $z=0.5$.  We take advantage of the remarkably wide modulation of electronic properties brought about by the Y/Eu substitution to highlight the net effect of the well characterized quenched disorder on the native clean limit cuprate.  By carefully calibrating the active layer hole content by several independent experiments, based on the Seebeck effect, \cite{CooperPRB1987, ObertelliPRB1992, HonmaPRB2004, SannaPRB2008,ConeriPRB2010} on the \msr\ internal fields in the AF phase\cite{ConeriPRB2010}, and on the $^{63}$Cu NQR measurement of chain length, we determine quantitatively controlled disorder, plane hole doping and their joint influence on the electronic properties. This is crucial to determine the plane hole number $h(z)$ (per square Cu(2)O$_2$), one of the two natural independent variables, together with the standard deviation $\sigma(z)$ for quenched disorder, against which we wish to assess the dependence of the electronic properties. By comparison with clean limit \ybcoy\ this provides the pure effect of quenched disorder in a crucial region of the phase diagram.

The paper is organized as follows: in Sec.~\ref{sec:Samples} we describe the \yebco\ sample preparation, their XRD resistivity and Seebeck  characterization which provides the hole count in the $z=1$ compound; Sec.~\ref{sec:DFT} reports DFT calculations and their main conclusions;  susceptibility and \msr\ measurements are presented in Sec.~\ref{sec:mSRSQUID}, with the description of strongly-underdoped \ybcoy\ hole count from \msr\ data; Sec.~\ref{sec:NQR} reports our NQR measurements, providing a determination of the average chain length, $\ell(z)$, and draws our conclusions on an absolute calibration of the hole count vs. $z$; Sec.~\ref{sec:Discussion} contains the main results, the phase diagram for \yeubco, quantifying the reduction in the ordering temperatures by structural disorder.

\section{\label{sec:Samples} Sample preparation and characterization}

A series of \yebco\ polycrystalline samples with increasing Yttrium content $z= 0.0, 0.15, 0.23, 0.3, 0.4, 0.6, 0.8$ and $1.0$ was prepared by the standard solid state reaction technique. The final oxygen content of the samples, $y=0.35(2)$, was achieved following a topotactic-like low temperature annealing\cite{MancaPRB2000,MancaPRB2001} of stoichiometric quantities of fully oxidized and fully reduced specimens ($y=0.98(2)$ and $y=0.05(2)$, respectively), tightly packed in vessels sealed in vacuum. The low temperature annealing allows the diffusion of the mobile species $O(1)$ inside the material towards the minimum energy configuration. The produced samples are thus chemically homogeneous, and the effective oxygen content was checked by thermogravimetry on each sample. Furthermore this procedure was shown to produce maximum length chain fragments at each composition and reproducible $y$ dependence of the electronic properties, coinciding\cite{ConeriPRB2010} with state of the art single crystal data.\cite{LiangPRB2006}

\subsection{\label{sec:XRD} Crystal structure}

X-ray diffraction (XRD) patterns were collected by using a Thermo Electron ARL X'tra powder diffractometer with Cu K$_\alpha$ radiation equipped with a Si(Li) solid state detector allowing to eliminate the incoherent contribution produced by the florescence of Eu. Data collections were performed with a step size of 0.03 degrees and a counting time of 2 sec per step.

Synchrotron X-ray diffraction data were collected for the samples with $z$ = 0.20, 0.40, 0.85, 1 on the ID15 beamline at the ESRF in Grenoble (France) by a Perkin-Elmer 2D CCD detector. The incident beam wavelength was set to 0.1423 \AA\ and the 2D images with 0.0083 degree per pixel were integrated by using the FIT2D software\cite{Hammersley1996} 
after instrumental parameters calibration with standard Ce$_2$O$_3$. LeBail fit of the patterns was carried out by using the GSAS software\cite{GSAS}
with EXPGUI interface\cite{EXPGUI}.

\begin{figure}
\includegraphics[trim=  0 0 0 0 , clip, width=0.4\textwidth,angle=0]{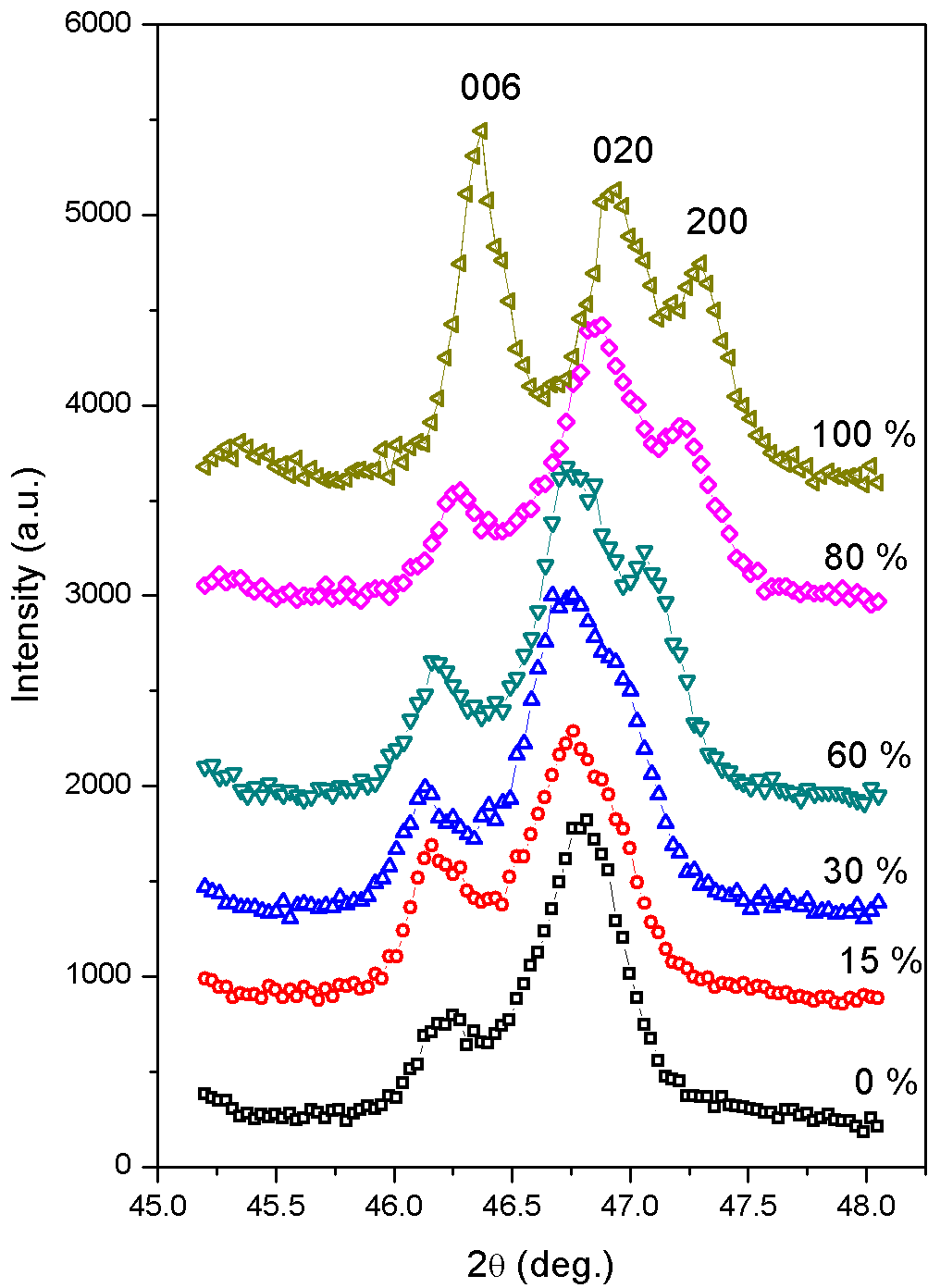}
\caption{\label{fig:XRD_patterns} (color on-line) Selected interval of the XRD patterns collected with $\lambda = 1.5406$ \AA. Percentages indicate the yttrium content.}
\end{figure}

We collected powder X-ray diffraction data for all the samples presented in this work. The evolution of a selected range of the diffraction patterns is reported in Fig.~\ref{fig:XRD_patterns}, clearly showing the splitting of the 200/020 reflection of the tetragonal cell for Y content higher than 50\%. Relevant microstructural effects giving rise in all cases to both asymmetric and hkl-dependent broadening of the peaks were detected, making the refinement of all the structural parameters through Rietveld analysis complex. Therefore,	 a thorough study of the lattice parameters as a function of the yttrium content was performed by LeBail fitting of the patterns. Two structural models, tetragonal (T) P4/mmm and (O) orthorhombic Pmmm, corresponding to the two end members of the series, $z=0,1$, respectively, were used to fit the data. 

\begin{figure}
\includegraphics[trim=  0 0 0 0 , clip, width=0.45\textwidth,angle=0]{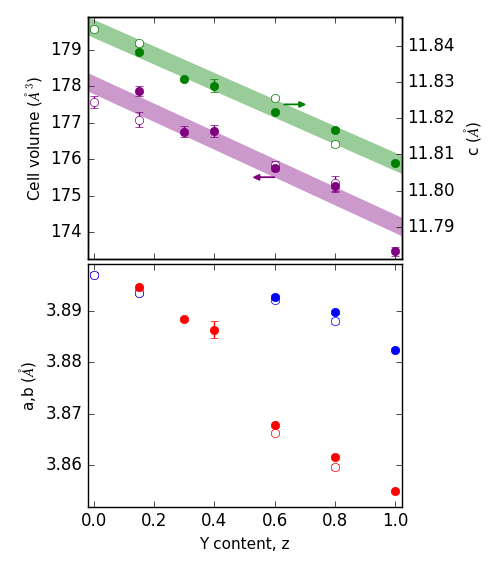}
\caption{\label{fig:XRD_volume} (color on-line) Cell volume and lattice parameters (the shaded lines are a guide to the eye). Open symbols: synchrotron data, filled symbols: laboratory data.}
\end{figure}

Figure \ref{fig:XRD_volume} displays the evolution of the lattice parameters and of the cell volume with $z$ (Y content), as determined by synchrotron and home based X Ray Diffraction (XRD). The linear scaling of both the cell volume and $c,(a+b)/2$ with $z$ is in agreement with the Vegard law, confirming the proper Eu/Y solubility throughout the full compositional range. The occurrence of the T-O phase transition at $z_c=0.5\pm 0.09$ indicates that the probability of chain formation vanishes along $a$ and becomes unitary along $b$ for $z<z_c$, whereas both probabilities are equal to 0.5 for $z<z_c$.

\subsection{\label{sec:RhoS} Transport properties, hole delocalization and quenched disorder}

In order to discuss the influence of quenched disorder on the electronic properties of our \yebco\ samples two key parameters are required: the carrier number $h$ per formula unit, per active Cu(2)O$_2$ layer, and an appropriate quantification of the degree of disorder. 

Let us start from the hole carrier number. Stoichiometry imposes that extra holes exist, transferred from core charges of nominal valence cations to other bands, mostly in the plane or in the chain layer. In \ybcoy\ band calculations\cite{FilippettiPRB2008, LopezPRB2010} predict a conductive band  on the Cu(1)O$_y$ sublattice for infinite long chains, consistent with experimental results for high oxygen doping $y>0.8$, where long chain lengths are expected\cite{BernhardPRB1995}. It is therefore safe to assume that in the Y compounds, for the present low $y$ values, $h$ accounts for all delocalized charges (with vanishing chain contribution). 

\begin{figure}
\includegraphics[trim=  0 0 0 0, clip, width=0.47\textwidth,angle=0]{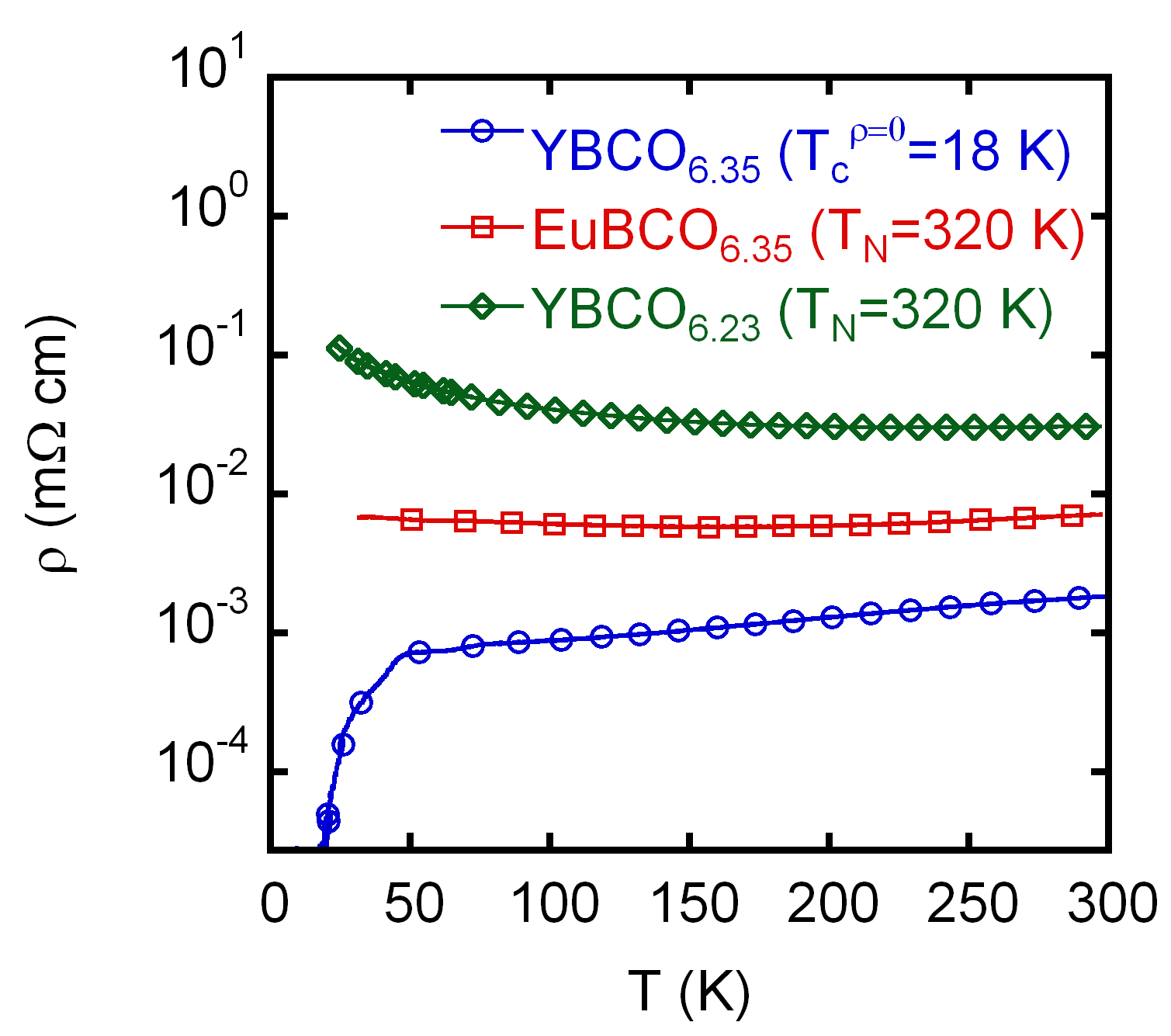}
\caption{\label{fig:rho} (color on-line) Comparison of the temperature dependence of electrical resistivity $\rho(T)$ for three selected samples of \yeubco, two with the same oxygen content $y$=0.35 but different electronic properties and two with different oxygen content but the same $T_N$=320 K. }
\end{figure}

Resistivity $\rho(T)$ measurements, performed by the standard four point technique on 2x2x14 mm$^3$ sintered specimens are shown in Fig.~\ref{fig:rho} for three specific samples: the end members \ybco\ and \eubco\ of the present series,  plus a \ybcol\ specimen (the same as in Ref.~\onlinecite{SannaSSC2003, SannaPRL2004, ConeriPRB2010}), chosen to have the same $T_N \approx320$ K as the first (Y) sample. The first and the last (Y) samples  display insulating and superconducting properties respectively. The Eu sample lies in between, still presenting insulating character down to 50 K. 
\ybcol\ and \eubco\ show rather different electric behavior, despite their very similar $T_N$ values. This indicates that for the Eu sample a fraction of mobile holes are present outside the planes already at this low oxygen content. Our data indicate that for this compound some conducting properties are associated with the disordered Cu(1)O$_y$ layer also for low oxygen content, still in the tetragonal phase, in contrast with \ybcoy\ case discussed above.

The room temperature Seebeck coefficient $S$ was demonstrated\cite{CooperPRB1987, ObertelliPRB1992, HonmaPRB2004, SannaPRB2008,ConeriPRB2010} to be rescalable to $h$ for the \ybcoy\ case. In a previous work\cite{SannaPRB2008} we accurately calibrated the exponential dependence of $S$ on the mobile holes and we further showed\cite{ConeriPRB2010} the good agreement with an independent calibration from the \textbf{c} lattice parameter\cite{LiangPRB2006} on oxygen doped \ybcoy single crystals. In sharp contrast with the wide modulation of electronic properties the  Seebeck data for \yeubco, shown in Table \ref{tab:Seebeck}, line up around two values, with a sharp jump around $z_c$, i.e. at the T-O phase transition. This may be understood assuming that the  Seebeck coefficient is determined by all mobile charges and with  coefficients strongly sensitive to crystal structure. In other words our empirical Seebeck calibration\cite{SannaPRB2008} is strictly valid only for the Y end member, but it cannot be used to measure the plane hole number $h$ for $z< 1$.

\begin{table}
\caption{Seebeck coefficient $S$ in $\mu$V/K for \yeubco\ and derived hole content, $h_{tot}$, using the \ybcoy\ calibration of Ref.~\onlinecite{SannaPRB2008}.}
\label{tab:Seebeck}
\centering\begin{tabular}{ccccccccc}
\hline
  z      &  1.0 & 0.8& 0.6 & 0.4 \\
\hline
$S$    &  82 & 82  & 91   & 157      \\
$h_{tot}$   & 0.071(2)  & 0.071(3) & 0.067(2) & 0.045(2) \\

\hline
z & 0.3 & 0.23 & 0.15 & 0.0 \\
\hline
$S$ & 131 & 152 & 158 & 141\\
$h_{tot}$ & 0.052(2)& 0.045(2) & 0.045(3)& 0.049(3)\\
\hline\end{tabular}
\end{table}

In order to determine the $z$ dependent additional contribution to mobile holes from the buffer layer we set up a formula unit calculation, that ignores band details, distinguishing only between $h$, and the remaining holes nominally assigned to the buffer layer. We assume a linear relation between Y content $z$ and the average Cu(1) ion charge $1+\lambda_0+\lambda_1 z$, to yield

\begin{equation}
\label{eq:h} 
Q_{tot}  = 2h(z)+\lambda_0+\lambda_1 z-2y = 0
\end{equation}

At fixed $y$ this linear relation may by understood in terms of a transfer function proportional to the distortion introduced by the larger cation, linear in $z$ according to the XRD data of Fig.~\ref{fig:XRD_volume}. Transfer is also proportional to the average chain fragment length, assuming a narrow distribution. This assumption will be discussed and justified in Sec.~\ref{sec:NQR} where we present the NQR derivation of the chain length dependence on $z$.

For $y=0.35$ we rely on the Seebeck calibration of the $z=1$ end member, $h_1=0.0700(25)$, to constrain Eq. \ref{eq:h}, obtaining $\lambda_0=0.56$. Another calibration is required in the AF region and it is provided by the \msr\ data of Sec.~\ref{sec:mSRSQUID}, and discussed in Sec.~\ref{sec:holes}.

Let us finally quantify quenched disorder. The agreement of the XRD data with the Vegard law indicates ideal randomness of the Y/Eu sublattice. The probability that a site is occupied by Eu is $z$, hence in a Poisson distribution its variance, $\sigma^2=z(1-z)$, provides an ideal case measure of quenched cation disorder.

\section{\label{sec:DFT} Ab initio calculations on structure and magnetism}

To clarify the relation between crystal structure, local structure and electronic properties,
we conducted a density-functional theory (DFT) investigation of four selected Y-Eu ratios
at the oxygen composition $y=0.325\simeq 1/3$ that best approximates the
experimental conditions. (We use the VASP code\cite{vasp} and GGA+U with the same
technicalities as in Ref. \onlinecite{lopez2010}, in particular U is tuned to get the correct gap in the undoped end member.)
\begin{figure}
\includegraphics[trim=  0 0 0 0 , clip, width=0.6\textwidth,angle=0]{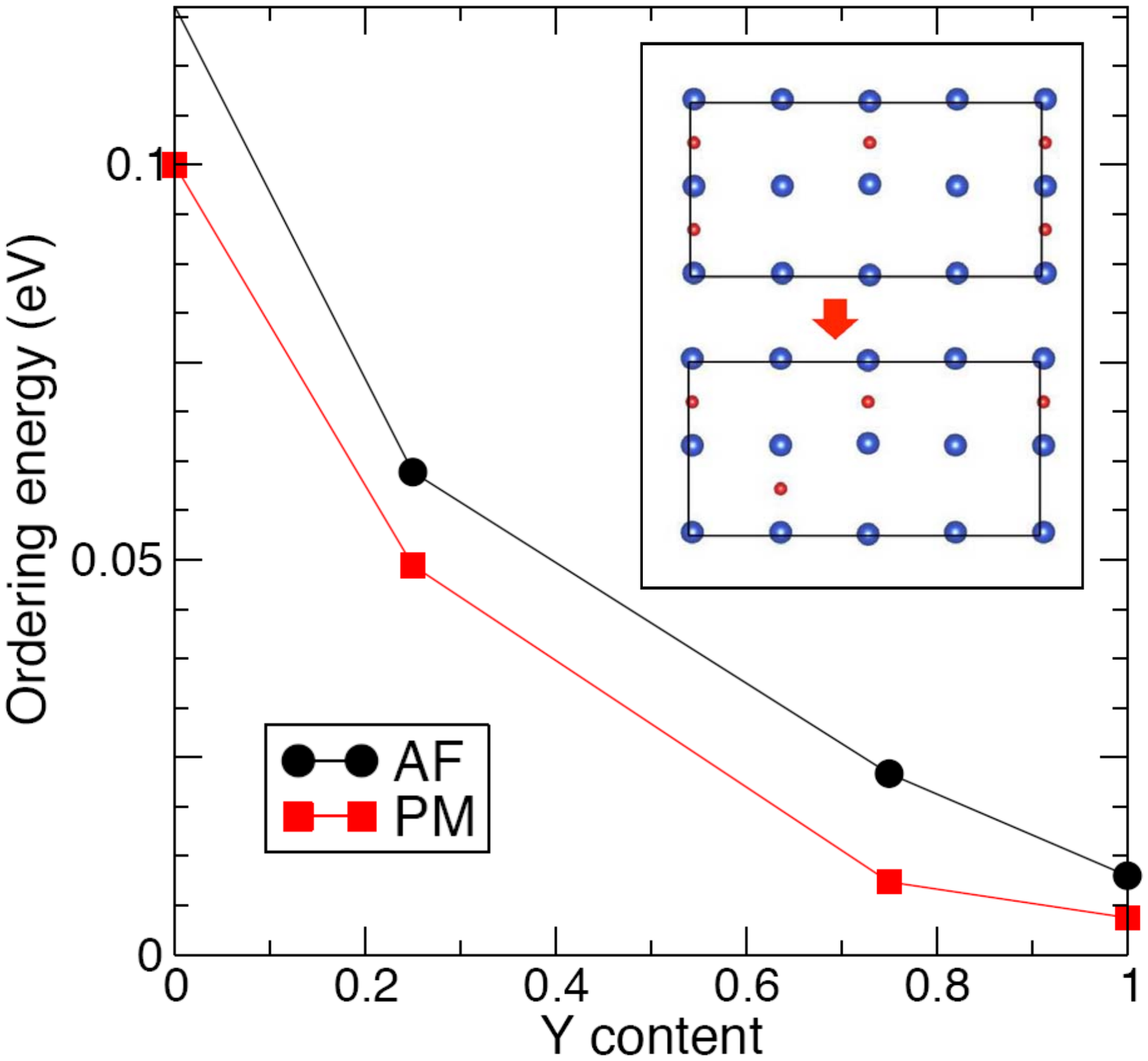} 
\caption{\label{fig:dE} DFT energy gain upon chain breaking as function of Y content in the AF-G and PM phase. Inset: Schematic of the chain-breaking transformation.}
\end{figure}

\begin{figure}
\includegraphics[trim=  0 0 0 0 , clip, width=0.45\textwidth,angle=0]{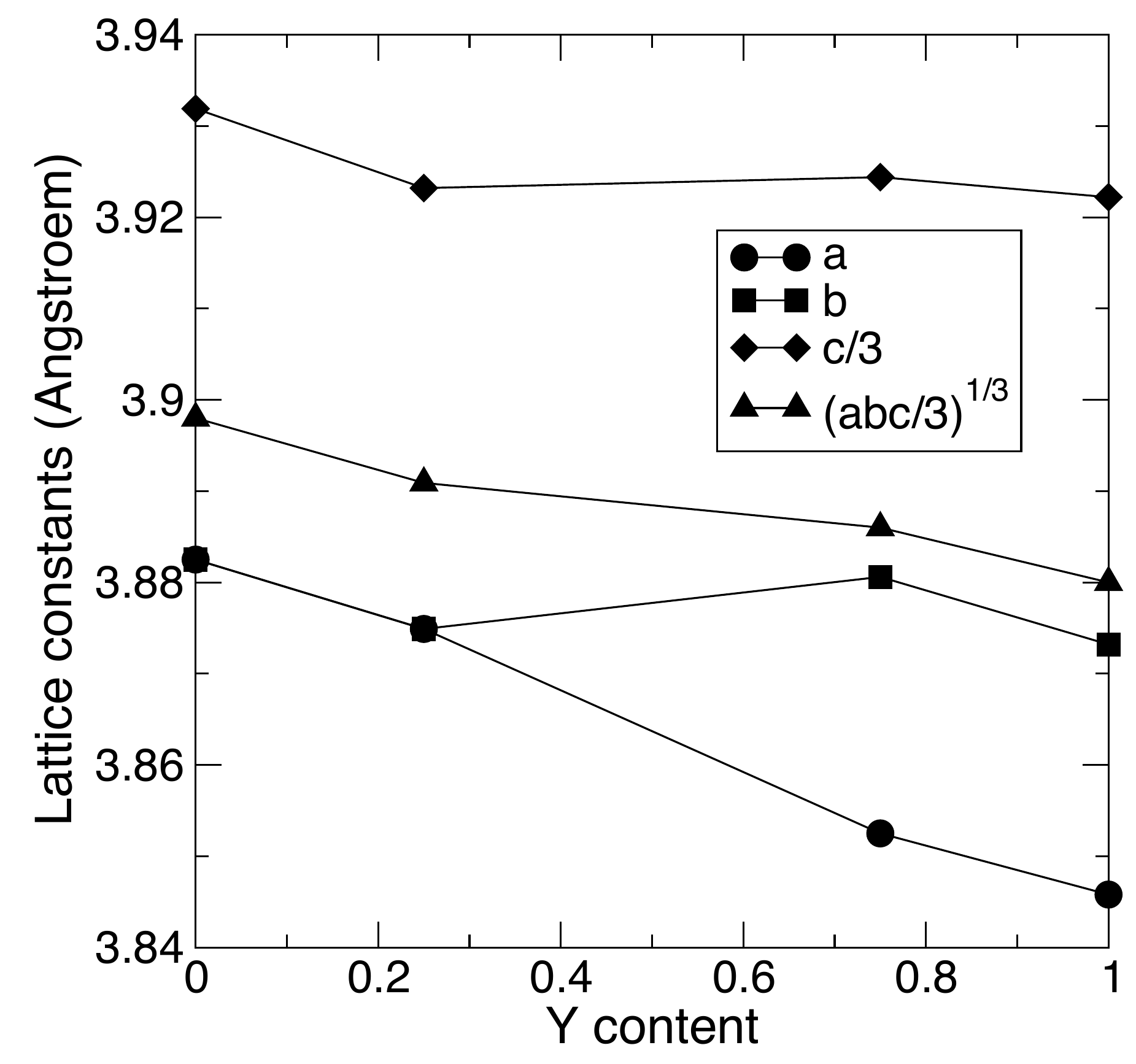} 
\caption{\label{fig:thlattice} Calculated cell volume and lattice parameters as a function of yttrium content (lines are guides to the eye).}
\end{figure}

The calculations reproduce several experimental features, notably the evolution of crystal
structure and ordering with composition, the approximate location of the structural phase
transition, an the hole content in the Cu planes. Figure \ref{fig:dE} displays the energy difference of infinite-chain and broken-chain structures (represented by the oxygen configurations in
the basal plane in the inset) vs.~composition. The main result (aside from the small
difference between paramagnetic and antiferromagnetic) is a clear energy gain for the
broken-chain configuration at the Eu end, that smoothly decays to a vanishing gain at the
Y end. This result backs up our understanding that the chain length (hence its influence)
vary smoothly in this system. In Fig.~\ref{fig:thlattice} we report the calculated values of $a, b, c$, and the average of all lattice constants for $z=0, 0.25, 0.75, 1$. The results are quite in line with experiment: the cell-volume reduction is well reproduced;  the $c$-axis length may indeed
be weakly plateauing at intermediate Y concentrations;  importantly, a T-O phase transition is predicted  somewhere around  50\% Y content.

\begin{figure}
\includegraphics[width=0.45\textwidth]{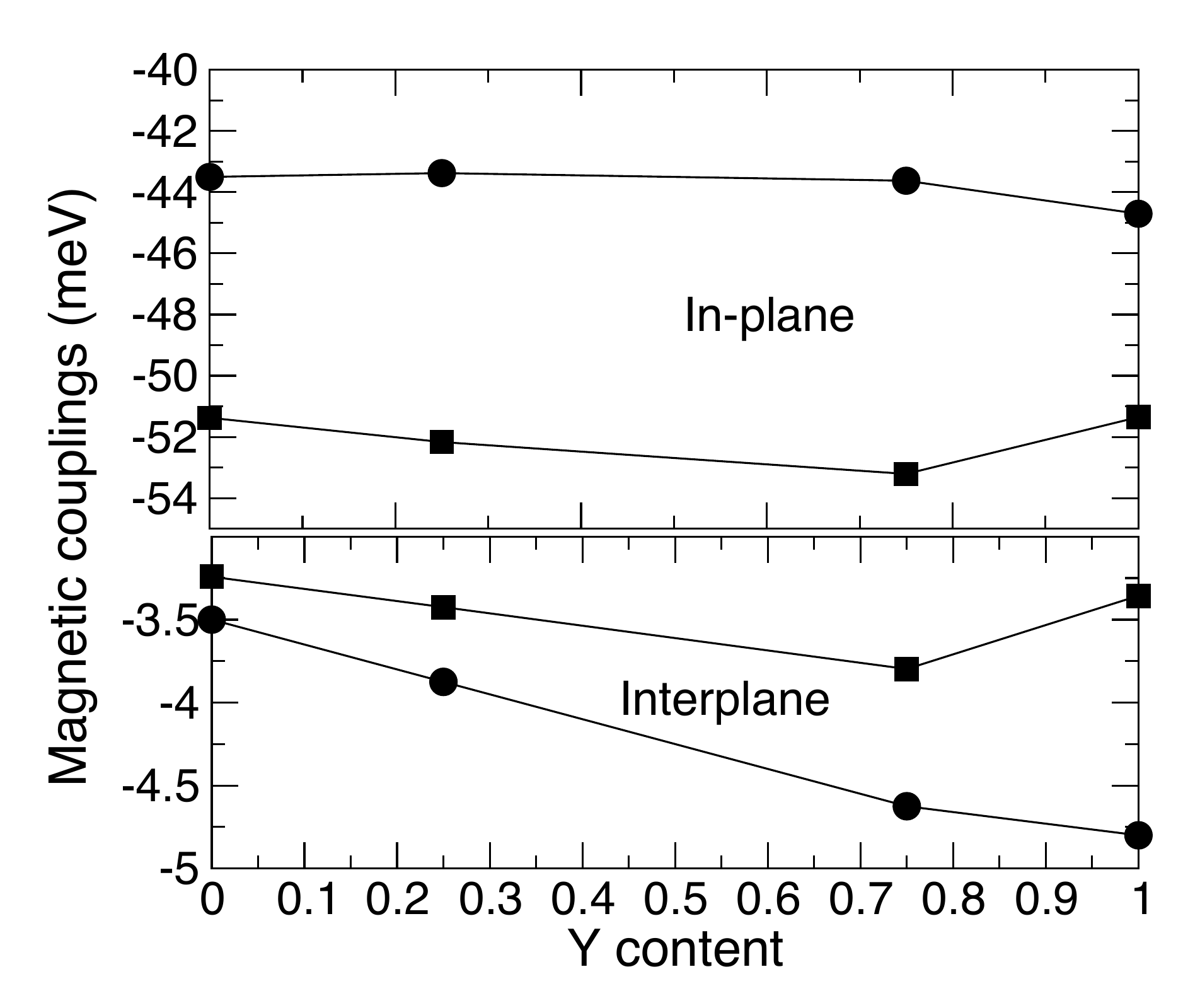}
\caption{\label{fig:J} Intra-plane (top) and inter-plane (bottom) magnetic couplings vs.~Y content for continuous (circles) and broken chains (squares).}
\end{figure}

We now make contact with the expectation that magnetism should disappear (albeit progressively, and not entirely) across the superconducting transition at $z_c\sim0.5$, with the antiferromagnetic (AF) ordered and insulating state giving way to a metallic, zero-moment state. The intra-plane and inter-plane first-neighbor Cu-Cu magnetic couplings vs.~Y concentration, extracted from the energies of various magnetic states (ferromagnetic, AF-G, AF-A) of the Cu sublattice, are
shown in Fig.\ref{fig:J}
to be roughly constant vs.~composition and  structure (they are also negative, indicating that
AF-G is the most stable of the non-zero-moment structures). Clearly this means that
T$_{N}$ would not change if the Cu moments remained unchanged throughout the
series; put differently,  AF-G will not disorder towards a moments paramagnet. What will
happen instead is that a zero-moment paramagnet (PM) takes over at some
concentration (as it does\cite{filippetti2010,lopez2010} in \ybcoy\ around $y$=0.3  ).
Indeed, the energetics shows that the PM state is favored over the AF at $z$=1 (\ybcod) and viceversa the AF is favored over the PM at $z$=0 (\eubcod).

In this picture, magnetic order weakens, and eventually disappears, due to the progressive
reduction, and eventual extinction, of Cu moments. Previous work\cite{filippetti2010, lopez2010}
(see in particular Fig.~14 of Ref.~\onlinecite{lopez2010}) suggested that this occurrence, and the
prevalence of the PM over the AF, correlates with increased hole transfer from the chains
to the planes: specifically, the AF state transfers less holes to the planes than the
paramagnetic zero-moments state.
Indeed, population analysis in the in-plane Cu $d$ states in the present case of Y substitution
shows again that a lower amount of holes is transferred to the plane in the Eu-rich than in the
Y-rich limit, the maximum difference between end compositions being of order 0.04/Cu.
This compares well with the  measured (or modeled) concentration change of mobile in-plane holes of Sec.~\ref{sec:holes}.
While correlation is not causation, this is circumstantial evidence that the hole transfer produced by Y substitution is key to the transition from the AF insulating state to a metallic
non-magnetic state.

\section{\label{sec:mSRSQUID} The electronic state: superconducting and magnetic properties}

\subsection{\label{sec:SQUID} Susceptibility}
\begin{figure}
\includegraphics[trim=  100 90 360 120, clip, width=0.45\textwidth,angle=0]{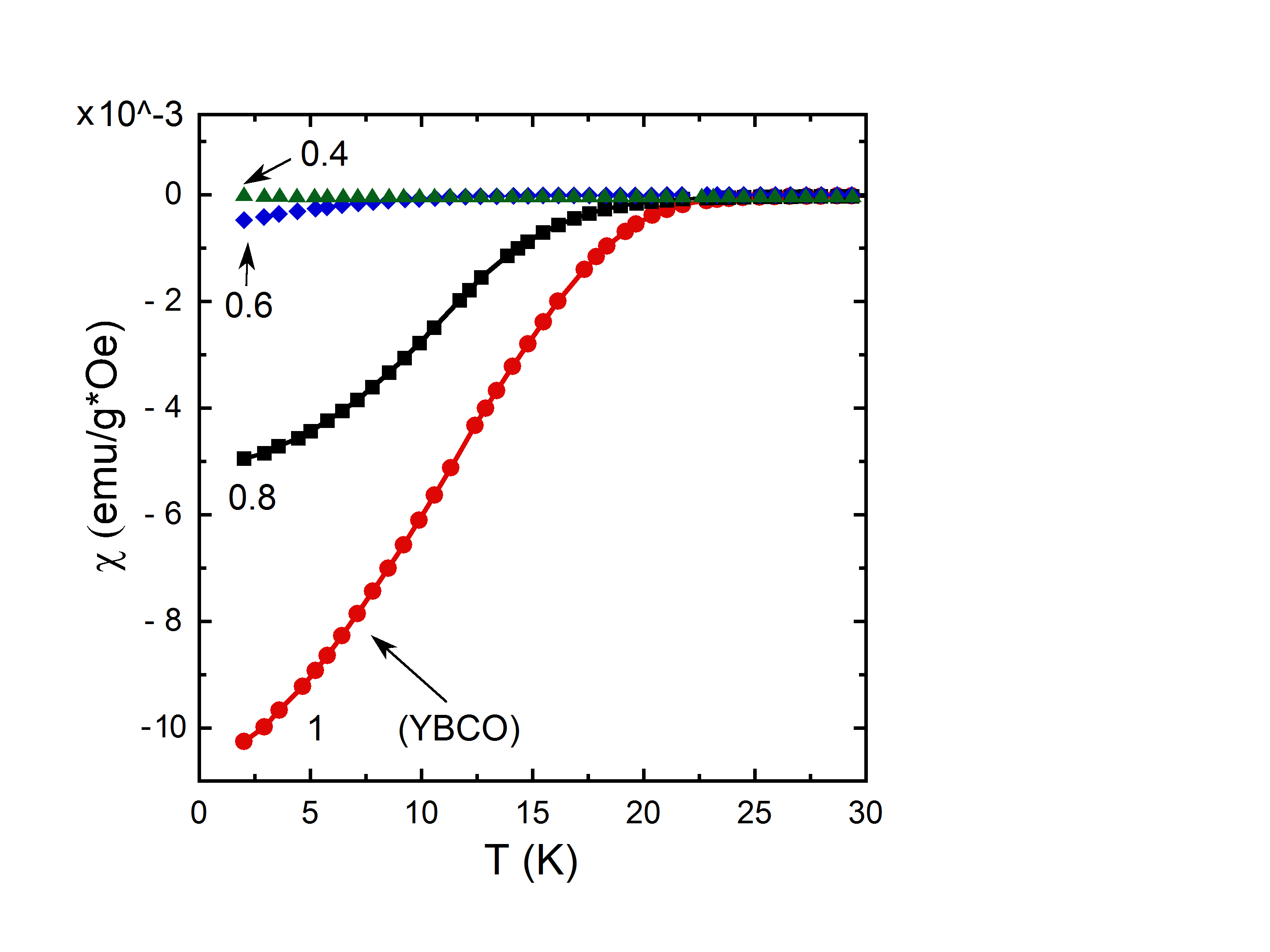}
\caption{\label{fig:Tc_SQUID} (color on-line) Zero-Field-Cooling (ZFC) susceptibility as a function of temperature with an applied field $H=0.2\unit{mT}$ of \yebco samples for different Yttrium content $z$. The substitution of Eu for Y progressively suppresses superconductivity.}
\end{figure}

Susceptibility was determined by measuring the sample magnetization in a field of $H =0.2$ mT in a MPMS system by Quantum Design Superconducting Quantum Interference Device (SQUID).
Figure \ref{fig:Tc_SQUID} shows the bulk Zero-Field-Cooling (ZFC) low temperature magnetic susceptibility of selected samples. A linear fit may be imposed on the 90\% to 10\% susceptibility drop, and we determine $T_c$ as the temperature where this line intercepts zero.
No traces of superconductivity have been detected up to $z=0.4$ (green triangles). A tiny diamagnetic shielding appears for $z=0.6$ (blue diamonds), becoming more robust for higher $z$ (higher Y content).

\subsection{\label{sec:MuSR} Zero Field \msr}

\begin{figure}
\includegraphics[width=0.45\textwidth,angle=0]{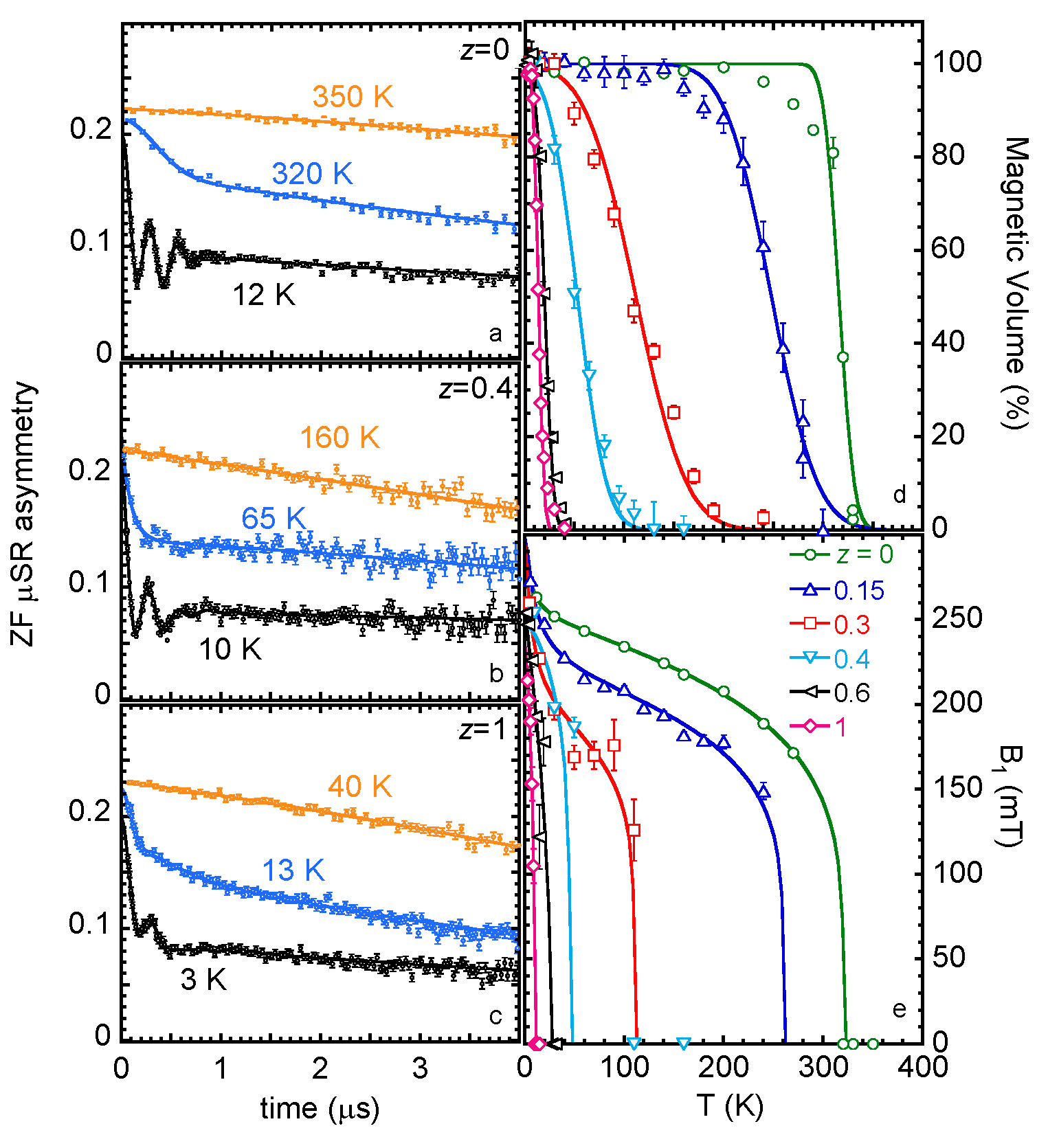}
\caption{\label{fig:ZFmuSR} (color on-line) Panels a), b) and c) display representative ZF-$\mu$SR time spectra for \yebco\ with $z=0,0.4,1$, respectively, with their best fit (see text). Panels d) and e) show the magnetic volume fraction $v_m$(T) and the internal field B$_1$(T) for a representative set of compositions, $z=0, 0.15, 0.3, 0.4, 0.6, 1$, as a function of temperature; the solid curves are the best fit following a $erf(T-T_N)$ function (d) and a power-law behavior (e) following Ref.\onlinecite{ConeriPRB2010} and \onlinecite{SannaPRBR2010} (see text).}
\end{figure}

Here we describe the $\mu$SR results and their consequences on hole calibration.
In a Zero Field (ZF) $\mu$SR experiment on powders the onset of antiferromagnetic order is directly detected from the appearance of a spontaneous local magnetic field $B_i$ at each inequivalent the muon site $i$, yielding a detectable muon spin precession oscillation pattern in the measured muon asymmetry. Muons bind to oxygen in two distinct sites, $i=1,2$ , chain-oxygen and apical-oxygen sites. \cite{ConeriPRB2010} A set of ZF-$\mu$SR asymmetry time spectra is displayed in the left panels of Fig.~\ref{fig:ZFmuSR} for three representative samples.  At the lowest temperature all the spectra display the presence of oscillations. Notably, also the samples $z=0.6,1.0$, bulk superconductor according to Fig~\ref{fig:Tc_SQUID}, show coexistence with low temperature magnetism (Fig.~\ref{fig:ZFmuSR} c). The $\mu$SR spectra have been analyzed following  Ref.\onlinecite{ConeriPRB2010}: well below the magnetic transition the best fit of the muon asymmetry  (solid curve) is given by 
\begin{equation}
A(t) = A_0 \left[\sum_i f_{T,i} e^{-(\sigma_i^2 t^2)/2 } \cos 2\pi \gamma B_it + f_L e^{-{\lambda t}}\right]
\label{eq:lowT}
\end{equation}
where $\gamma=135.5 $ MHz/T is the muon gyromagnetic ratio.  The transverse and longitudinal muon weights $f_{T,i}, f_L$ account for muon spin components  perpendicular and parallel to the local field directions, experiencing relaxation rates $\sigma_i$ and $\lambda$, respectively,  and obey $\sum_i f_{T,1}+f_L=1$. Magnetic transitions are inhomogeneous and $v_m=3(1-f_L)/2=3(f_{T,1}+f_{T,2})/2$ measures the magnetic volume fraction (Fig.~\ref{fig:ZFmuSR}d).

The intermediate temperature in each panel of Fig.~\ref{fig:ZFmuSR}a,b,c shows data just below $T_N$, characterized by over-damped precessions and increased value of the longitudinal fraction $f_L$. At the highest temperatures, well above the magnetic transition, samples are fully in the paramagnetic state and the asymmetry is fitted to the function $A(t) = A_0 e^{-({\sigma_n}^2 t^2)/2 }$
where $\sigma_n\approx0.2 \mu s^{-1}$ is the low static relaxation rate arising from the very much smaller neighboring nuclear moments. 

Figure \ref{fig:ZFmuSR} d,e display the temperature evolution of a representative set of Eu/Y compositions of the magnetic volume fraction $v_m$(T) and the mean field $B_1(T)$, proportional to the staggered magnetization. \cite{ConeriPRB2010} The N\'eel transition follows a narrow Gaussian distribution, its mean being the temperature where $v_m(T_N)=0.5$, very close to where $B_1$ vanishes.
For $z=0.4$ and 0.6 the magnetic transitions temperatures are $40(20)$ K and $20(3)$ K respectively. Therefore in the range $0.4<1-z<0.6$ the very sharp drop of $T_N(h)$ flattens abruptly, turning into the low spin freezing temperature $T_f$ characteristic of the cluster spin glass state\cite{JulienPRL1999}, in close analogy with \lsco\ and \ycabcoy. 

Therefore SQUID susceptometry and \msr\ provide clear evidence that indeed \yebco\  spans a wide portion of the typical cuprate phase diagram, displaying a clear deviation from the carefully characterized behavior\cite{ConeriPRB2010} of {\em clean} \ybcoy, in that it shows an intermediate region of where neither the long range AF state nor the superconducting state are detected. This is the hallmark of quenched disorder.\cite{AlvarezPRB2005,DagottoScience2005}

In the following we assume that the N\'eel temperature $T_N(z)$ and the local field $B_\mu(z,T)$ are largely insensitive to the rare earth at the Y site, as it is the case in the related system \calabalacuo.\cite{KerenPRB2003} As it was shown in different cuprate families by $\mu$SR\cite{NiedermayerPRL1998,SannaPRL2004,SannaSSC2003,ConeriPhysicaB2009,ConeriPRB2010} and by $^{139}$La NQR in \lsco, \cite{BorsaPRB1995} the temperature evolution of the internal field $B_i$ (Fig.~\ref{fig:ZFmuSR}e) displays a marked deviation from a simple magnetic order parameter behavior, in the form of a characteristic upturn at low temperatures.  The function $B_i(T)$ was described by a phenomenological model\cite{ConeriPRB2010} of thermally activated holes, whose number is identified with $h$:  

\begin{equation}
\label{eq:fitB}
B_1(h,T)= B_F \left[ 1-\frac h {h_c} e^{-T_A/T}\right]^\alpha (1-T/T_N)^\beta
\end{equation}

where $B_F=30$ mT is the value of the local field at muon site 1 for $T\rightarrow 0$ and $h=0$ (i.e. in the parent compound), exponents values are $\alpha\approx0.3$, $\beta\sim0.15-0.2$, and fixed the critical hole number is $h_c=0.056(2)$, in agreement with the direct determination from $T_N(h)$.  

In this model $T_A$ is a crossover temperature between a high temperature ($T\gg T_A$) thermally activated regime, with plane mobile holes $h$, and a low temperature frozen regime ($T\ll T_A$), clearly visible as an upturn\cite{BorsaPRB1995, ConeriPRB2010, SannaPRBR2010} also in Fig.~\ref{fig:ZFmuSR}e, where the holes localize. 

Our model mimics  the effect of the thermal excitations of the charge doped 2D Heisenberg quantum square antiferromagnet (2DHQSAF)\cite{Capati2015}. Hole activation is also reflected in a peak in the longitudinal relaxations $\lambda$ in the same range of temperatures (not shown), indicating a related spin-activation process\cite{NiedermayerPRL1998,SannaSSC2003}. Finally, when the normalized correlation between parameters $T_N$ and $T_A$ exceeds 0.9 we fix the two to the same value and call it freezing temperature $T_f$. Equation \ref{eq:fitB} agrees perfectly with the present set of data, as show by the solid best fit curves of Fig.~\ref{fig:ZFmuSR}e. 

Plane hole number may be thus determined from the known\cite{ConeriPRB2010} dependence of $T_N$ on $h$ in \ybcoy 

\begin{equation}
\label{eq:fitTN}
T_N(h)=T_{N0}\left[1-\left(\frac h{h_c}\right)^2\right]
\end{equation}

with $T_{N0}=422(5)$ K and $h_c=0.056(2)$. The best fit of the $z=0$ sample data, Fig.~\ref{fig:ZFmuSR}e, to Eq.~\ref{eq:fitB} yields $h=0.026(2)$ and $T_N=320(15)$ K. The latter, inserted into Eq.~\ref{eq:fitTN} provides a value of $h=0.027(2)$, in perfect agreement with the best fit value.
Remarkably, this method allows us to compute the plane hole number directly, since the magnetic properties are strictly influenced by local charge doping, in contrast to both  transport measurements, sensitive to all mobile charges, and structural measurements, globally reflecting the changes in all layers.

Let us finally recall that the superconducting transition temperatures of the clean limit \ybcoy\ are related\cite{LiangPRB2006} to $0.05\le h\le h_{opt}=0.16$ by 

\begin{equation}
\label{eq:fitTc}
T_c(h) = 82.6 (h_{opt} - h)^2
\end{equation}

\section{\label{sec:NQR} Nuclear Quadrupole Resonance, chain length and plane hole count}
In \ybcoy\ the hole carrier number $h$ depends on both $y$ and sample preparation conditions. Reproducible values are obtained after very long annealing slightly above room temperature. \cite{JorgensenPC1990,VealPRB1990,VealPC1991} In annealed samples, like the ones investigated here, the average chain length $\ell$, defined as the average number of oxygen ions in Cu(1)O chain fragments, depends only on oxygen content ($y$): filled Cu(1)O chains run along the orthorhombic crystal $b$ direction for $y\approx 1$, and for decreasing oxygen content sequences of empty and filled chains tend to alternate regularly along $a$ forming well defined superstructure. Below a critical threshold ($y_c\approx 0.30$ for \ybcoy), known to depend on the average cation radius, short fragments of chains become orientationally disordered.  \cite{DeFontaineNature1990,Beyers1989,MancaPRB2000,MancaPRB2001,ZimmermannPRB2003} This condition marks the T-O phase transition and identifies a critical chain length $\ell_c$, also dependent on cation radius. 

\subsection{\label{sec:chainlength} Chain length determination}

\begin{figure}
\includegraphics[trim=  0 0 0 0, clip, width=0.40\textwidth,angle=0]{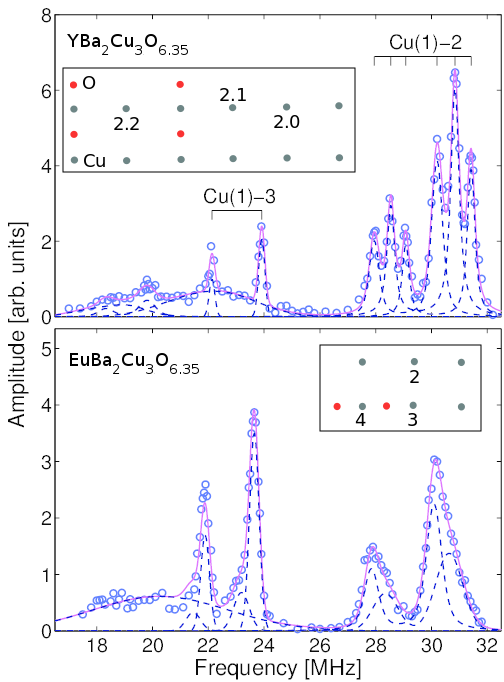}
\caption{\label{fig:NQR} (color on-line) NQR spectra with best fit curves at $T=77 \unit{K}$ for the two end members of the Y/Eu series. The Cu(1)-$i$ site $i=2,3,4$, distinguished by their nearest-neighbor, are shown in the bottom inset (see text); the top inset shows Cu(1)-2.$j$ sites, $j=0,1,2$ with three distinct next-nearest-neighbor environments, justifying the fine splitting of the Cu(1)-2 peaks.}
\end{figure}

The NQR resonance frequency of Cu, proportional to the electric field gradient at the nucleus, is characteristic of each inequivalent copper site in the lattice and very sensitive to nearest-neighbor (nn) oxygen configuration. Important nn configurations are Cu(1)-2 and Cu(1)-3, shown in the bottom inset of Fig.~\ref{fig:NQR}: the former is bonded only to two apical O ions, and it is surrounded by four O vacancies in the basal layer, the latter is bonded to an additional nn chain O, with three nn O vacancies. Since there are two Cu isotopes, $^{63}$Cu and $^{65}$Cu, each configuration gives rise to an isotope doublet, at the well known frequency ($\nu_{63}/\nu_{65}=1.082$) and intensity ($I_{63}/I_{65}=2.235$) ratios.

Two representative Cu(1) NQR spectra for the end members \ybco\ (A) and \eubco\ (B) are shown in Fig.~\ref{fig:NQR}. The spectra are fitted to a sum of Voigt functions. According to a well established line assignment\cite{LuetgemeierPC1996, MancaPRB2000,SannaSSC2003}, the narrow doublet peak at 24 MHz and 22 MHz is due to the Cu(1)-3 configuration. The large peaks at $\approx 22 \unit{MHz}$ is due to Cu(1)-4 sites from longer chain fragments.

Finer details are seen in the Y sample, where three distinct $^{63}$Cu resonance peaks in the $29-32 \unit{MHz}$ range and their very similar $^{65}$Cu isotope replicas in the 27-29 MHz range correspond to as many different next-nearest-neighbor configurations of Cu(1)-2. They are shown in the top inset of Fig.~\ref{fig:NQR}: Cu(1)-2.0 surrounded by empty chains on both sides (at 30.1 and 27.8 MHz), Cu(1)-2.1, with one empty and one filled chain (at 30.6 and 28.3 MHz) and Cu(1)-2.2 with filled chains on both sides (31.1  and 28.7 MHz). \cite{HeinmaApplMagnReson1992} The well resolved narrow resonances of the \ybco\ sample testify that the three different local configurations are themselves quite homogeneous, as expected from the relatively high degree of order. On the contrary, these three peaks get broader and unresolved for the \eubco\ sample, indicating a greater degree of local disorder in the basal layer. This is further supported by the fact that for this sample the Cu(1)-3/Cu(1)-2 spectral weight ratio increases, indicating a relative abundance of the chain-end sites, i.e. the substitution of Eu yields to a shortening of the chain length at fixed oxygen content. 

The chain length $\ell$, shown in Fig.~\ref{fig:Ortho.S.l}c, is calculated from the weight of the two configurations (the area under each NQR peak doublet)  as in Refs.~\onlinecite{LuetgemeierPC1996, MancaPRB2000} 

\begin{equation}
\label{eq:chain}
\ell=3/7 (2A_2 /A_3+1)
\end{equation}

\subsection{\label{sec:holes} Plane hole number calibration.}

In Sec.~\ref{sec:RhoS} we assumed a linear relation, Eq.~\ref{eq:h}, between plane hole number $h$ per Cu(2)O$_2$ square cell and the Eu content $z$. We wish now to discuss the basis of this assumption.

\begin{figure}
\includegraphics[trim=  0 0 0 0 , clip, width=0.4\textwidth,angle=0]{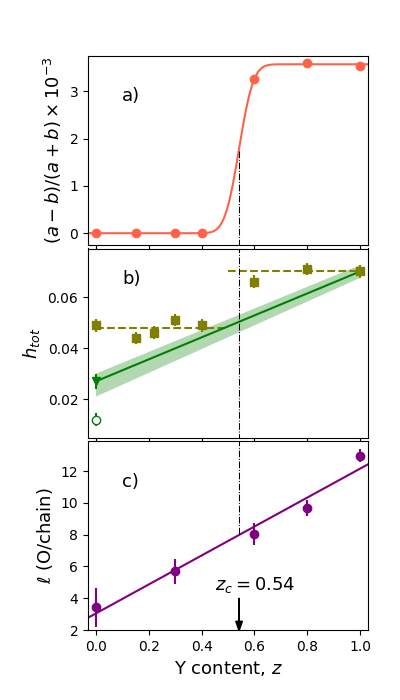}
\caption{\label{fig:Ortho.S.l} (color on-line) Dependence on Y content $z$ of a) orthorhombic distortion, the vertical dashed line indicates the T-O structural transition at $z_c$; b) total hole count $h_{tot}$ derived from Seebeck with \ybcoy\ calibration\cite{SannaPRB2008} (filled squares) and mobile plane holes $h$ for the end members, from Eq.~\ref{eq:fitB} (open circle) and from Eqs.~\ref{eq:fitTN} (filled triangle); the solid line with shaded band is the linear relation $h(z)$ with its estimated uncertainty; c) average chain length $\ell$, in O atoms per chain, from Eq.~\ref{eq:chain}.}
\end{figure}

Figure \ref{fig:Ortho.S.l} a,b shows the variation of the orthorhombic distortion coefficient $(a-b)/(a+b)$ and the total hole number $h_tot$, from the Seebeck data of Tab.~\ref{tab:Seebeck}, vs. the Y content $z$.  Both parameters are constant in the two distinct T and O structures and they both jump at the first order phase transition, for $z=z_c$. By comparison the chain length $\ell$ obtained by NQR with from Eq.~\ref{eq:chain} and shown in panel c varies linearly with  $z$ and is insensitive to the phase transition. 

The hole transfer mechanism from the basal layer onto the Cu(2)O$_2$ planes requires the oxidation of Cu(1) from Cu$^{1+}$ to Cu$^{2+}$, but it becomes active only above a minimal chain length $\ell_m\approx 2$,  \cite{Uimin1992,MancaPRB2000} after which both $\ell$ and $h$ increase. Heinma {\em et al.} \cite{HeinmaApplMagnReson1992} (see Fig.~5 therein) showed by NQR  in \yRbco\ that a linear dependence holds between $\ell$ and oxygen content $y$, with equal slope and slightly different intercepts in Re=Y, Gd.  The two linear relation result in a direct proportionality between $y$ and $h$, which was experimentally established in \ybcoy\  by the extensive single crystal work of Liang {\em et al.}, Ref.~ \onlinecite{LiangPRB2006} (notice that very accurate linearity holds in the underdoped regime relevant for the present work).

By the same argument  we conclude that that the linear relation between $\ell$ and $z$, shown in Fig.~\ref{fig:Ortho.S.l}c, implies a linear scaling of $h$ with $z$, i.e. the validity of Eq.~\ref{eq:h}.

Recalling that the \yebco\ series displays a linear dependence on $z$ for both the cell volume and the average lattice parameters, Fig.~\ref{fig:XRD_volume}, it is reasonable to conclude that the steric hindrance of the Eu cation forces the \ybco\ lattice to a nearly isotropic dilation proportional to Eu content, which in turns determines the average chain length.

\section{\label{sec:Discussion} Discussion and conclusion: disorder influence on the phase diagram.}

\begin{figure}
\includegraphics[trim=  0 0 0 0 , clip, width=0.45\textwidth,angle=0]{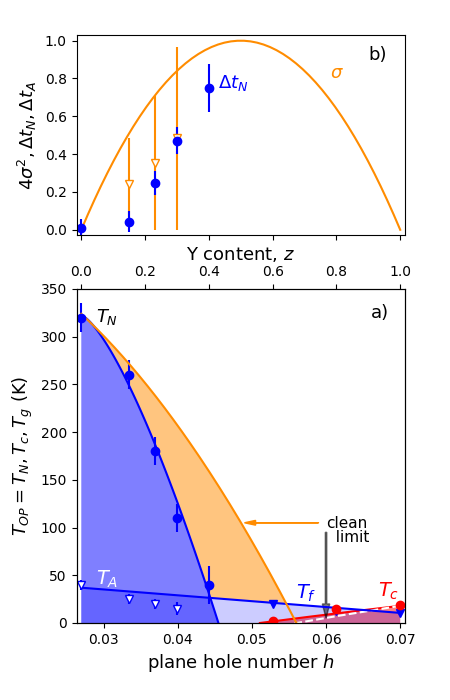} 
\caption{\label{fig:PhaseDiagram} (color on-line) a) Phase diagram of \yeubco\ as a function of plane hole number per CuO$_2$ square cell (bottom) and Y content (top); see the text for the lines through the data. b) Normalized disorder variance (see Sec.~\ref{sec:RhoS}), $\Delta t_N, \Delta t_A$ vs. Y content.}
\end{figure}

The main result of this work is summarized in Fig.~\ref{fig:PhaseDiagram}, where the  transition temperatures in \yebco\ are plotted versus hole concentration $h$, obtained according to the criteria of  Sec.~\ref{sec:Samples} and \ref{sec:mSRSQUID}. Blue filled circles, open and filled triangles represent best fit Ne\'el temperatures $T_N$, activation temperatures $T_A$ and freezing temperatures $T_f$, respectively, from the data of Fig.~\ref{fig:ZFmuSR}, by means of Eq.~\ref{eq:fitB}, while red filled circles are superconducting temperatures from SQUID measurements.

It is remarkable that, although the substitution parameter $z$ does not directly introduce a chemical doping, the phase diagram reproduces qualitatively the universal HTC cuprate behavior. A similar equivalence of  chemical doping and structural distortion is well known in iron pnictides  and selenides\cite{KimberNatMat2009,MitzuguchiSupercSciTech2010}, but is unprecedented in cuprates.

The influence of disorder is shown by the deviations from the \ybcoy\ nearly clean-limit case, that is represented by the continuous orange and dashed white lines in Fig.~\ref{fig:PhaseDiagram}a  (referred to $h$, bottom axis). The difference  is apparent when comparing the latter two curves with data (symbols) and their guide to the eye, delimiting the \yebco\ AF, cluster spin glass and superconducting phases, represented by the blue, light blue and pink shaded areas respectively. Three remarkable features stand out: {\em i)} a strong suppression of $T_N$, maximum around $z_M=0.5$, where disorder is maximum; {\em ii)} the appearance of a wide region where the cluster spin glass temperature $T_f$ is the largest energy scale;  {\em iii)} a surprising enhancement of $T_c$ (red solid circles) over the clean case (white dashed line).

We start by noticing that in cuprates values of $h$ of just few percent are sufficient to suppress $T_N$. This dramatic hole doping effect, much more effective than moment suppression by dilution\cite{Carretta2011}, is due to frustration in the N\'eel state of the 2DHQSAF, introduced by the spin-hole excitations. \cite{Capati2015} The rather strong additional reduction due to disorder (point {\em i)}) is surprising, since both magnetic moments on Cu and their exchange couplings are virtually unaffected by the Y-Eu substitution. This suggests that quenched disorder influences directly spin-hole excitation, possibly by pinning, thus introducing additional frustration. The  numerical work available to date\cite{DagottoScience2005}  does not allow a direct comparison with our data. This calls for further theoretical work on the excitations of the 2DHQSAF in the presence of extra charge carriers and disorder.

Point {\em ii)} confirms the progressive opening of a spin glass gap in the weak disorder regime, already observed for strong disorder in \ycabcoy. The onset spin glass temperatures $T_f$ of the $0.6\le z \le 1.0$ samples fall on a straight $T_A(h)$ line from the clean $z=0$ value of $T_A$, exactly like in Ref. \onlinecite{ConeriPRB2010,SannaPRBR2010}. The normalized suppression of $T_N$ and $T_A$, defined as $\Delta t_{N,A}=1-T_{N,A}(h,z)/T_{N,A}(h,0)$ are shown in Fig.~\ref{fig:PhaseDiagram}b.  They correlate with the normalized disorder variance $4\sigma^2$, shown in  the same plot. The correlation is non linear, since isovalent cation disorder is ineffective below 10\% substitution, as it was already demonstrated in Ref.~\onlinecite{ConeriPRB2010}. 


Let us finally consider point {\em iii)}, the shift of the onset concentration for superconductivity under increased quenched disorder. It is small, but outside errorbars and surprisingly negative: disorder seems to favour superconductivity. It is tempting to argue that the suppression of the 2DHQSAF provides an increased spin density of state involved in the superconducting pairing, but in \ycabcoy\ the same shift was comparable and positive, seemingly in contradiction with this argument. A tentative explanation is to consider that the onset of superconductivity actually coincides, within the accuracy of the present study, with the T-O transition as it does in pure \ybcoy. The locking of the $T_c$ onset and the structural transition emphasizes the role of nematic susceptibility\cite{Nie2014} in unconventional superconductivity.

\begin{acknowledgments}
We thank Edi Gilioli and Francesca Licci for useful discussions on sample preparation and characterization, Alex Amato and James Lord for support in the $\mu$SR experiments, We gratefully acknowledge the Science and Technology Facilities Council (STFC) and the Paul Scherrer Institut, Villigen, Switzerland  for access to their muon beamtime at ISIS and LMU. Work supported in part by UniCA, Fondazione di Sardegna, and Regione Autonoma della
Sardegna via Progetto biennale di ateneo 2016 Multiphysics approach to
thermoelectricity, by CINECA, Bologna, Italy through ISCRA Computing Grants, and by
CRS4 Computing Center, Piscina Manna, Italy.

\end{acknowledgments}

\bibliography{PRBEu}

\end{document}